\documentclass[3p,times,procedia]{elsarticle}
\usepackage{ecrc}
\usepackage[bookmarks=false]{hyperref}
    \hypersetup{colorlinks,
      linkcolor=blue,
      citecolor=blue,
      urlcolor=blue}

\volume{00}

\firstpage{1}

\journalname{Procedia Computer Science}

\runauth{Pedro Pereira}

\jid{procs}

\usepackage{amssymb}
\usepackage{glossaries}
\glsdisablehyper
\usepackage{booktabs}
\usepackage{tabularx}
\makeatletter
\AtBeginDocument{
  \let\oldciteauthor\citeauthor
  \renewcommand{\citeauthor}[1]{\hypersetup{hidelinks}\oldciteauthor{#1}}
}
\makeatother
\usepackage[figuresright]{rotating}
\usepackage{soul}
\usepackage{xcolor}
\definecolor{revieworange}{RGB}{255,126,39}
\newacronym{AI}{AI}{Artificial Intelligence}
\newacronym{NLP}{NLP}{Natural Language Processing}
\newacronym{LLMs}{LLMs}{Large Language Models}
\newacronym{RAG}{RAG}{Retrieval-Augmented Generation}
\newacronym{ASR}{ASR}{Attack Success Rate}
\newacronym{OLS}{OLS}{Ordinary Least Squares}

\begin{document}
\begin{frontmatter}



\dochead{30th International Conference on Knowledge-Based and Intelligent Information \& Engineering Systems (KES 2026)}%

\title{Influence Factors on RAG Poisoning}


\author[a]{Pedro Pereira\corref{cor1}} 
\author[a]{Eva Maia}
\author[a]{Isabel Praça}
\author[b]{Adrien Bécue}

\address[a]{GECAD, ISEP, Polytechnic of Porto, rua Dr. António Bernardino de Almeida, 4249-015 Porto, Portugal}
\address[b]{THALES SIX GTS, Gennevilliers, France}

\begin{abstract}
%Retrieval-Augmented Generation (RAG) systems enhance large language models by grounding responses in external knowledge, but they are also vulnerable to poisoning attacks in which adversarial documents manipulate retrieval and generation. 
Retrieval-Augmented Generation (RAG) systems enhance large language models by grounding responses in retrieved documents from external knowledge sources at inference time. However, this reliance on retrieved content introduces vulnerabilities to poisoning attacks, in which adversarial documents can manipulate both the retrieval process and the generated outputs. This paper investigates poisoning robustness in RAG through a full factorial experimental study covering 432 configurations. We analyze the impacts of dataset, retriever type, retrieval depth, database composition, chunking strategy, and generator model on retrieval-level and generation-level metrics. The results show that retriever architecture, dataset, and retrieval depth are the strongest factors affecting poisoning exposure, while generator choice and database composition have a major impact on downstream attack success. Dense and graph-based retrievers generally improve robustness relative to BM25, whereas larger retrieval depth increases the likelihood of retrieving poisoned passages. We further show that replicating poisoned content across multiple databases amplifies adversarial influence, while additional clean sources can mitigate it. These findings highlight that poisoning vulnerability in RAG is not attributable to a single component, but instead arises from the interaction of retrieval, generation, and knowledge-base configuration.
\end{abstract}

\begin{keyword}
Retrieval-Augmented Generation; poisoning attacks; adversarial information retrieval; large language models ;robustness; secure RAG

%% keywords here, in the form: keyword;keyword

%% PACS codes here, in the form: \PACS code;code

%% MSC codes here, in the form: \MSC code;code
%% or \MSC[2008] code;code (2000 is the default)

\end{keyword}
\cortext[cor1]{Corresponding author. Tel.: +351 22 83 40 532 ; fax: +351 22 83 21 159.}
\end{frontmatter}

%\correspondingauthor[*]{Corresponding author. Tel.: +0-000-000-0000 ; fax: +0-000-000-0000.}
\email{peesp@isep.ipp.pt}

% \linenumbers


%\enlargethispage{-7mm}
\section{Introduction}\label{sec:1}
Recent advances in \gls{AI}, particularly in \gls{NLP}, have been driven by the emergence of \gls{LLMs}, which have demonstrated strong capabilities in a wide range of tasks \cite{SajjadiMohammadabadi2025ASO}. Their rapid adoption in consumer and enterprise use cases has positioned \gls{LLMs} as a key technological component in applications such as question answering, summarization, coding assistance, customer support, and decision support \cite{Muhammad_2025}. However, LLMs remain constrained by two well-known limitations. First, their internal knowledge becomes outdated over time, which means that retraining and updating processes are constantly necessary \cite{Fan2024ASO}. Furthermore, they are prone to hallucinations because they function as probabilistic pattern matchers. Without access to the necessary knowledge, they prioritize linguistic fluency and logical consistency over factual accuracy, often generating results that seem plausible but are entirely fabricated \cite{Fan2024ASO}.

To mitigate these weaknesses, \gls{RAG} has emerged as a widely adopted architectural paradigm \cite{Lewis2020RetrievalAugmentedGF}. Instead of relying exclusively on parametric knowledge stored in model weights, \gls{RAG} augments the generation with external information retrieved from a knowledge base at inference time \cite{Lewis2020RetrievalAugmentedGF}. This design enables access to up-to-date and domain-specific information without retraining the model, while also improving factual foundations and contextual relevance \cite{ LI2025100417}. However, \gls{RAG} introduces a new and important security dependency, the reliability of the generated answer becomes directly related to the integrity of the retrieved external knowledge \cite{Zou_2025}. In an \gls{RAG} pipeline, the retrieval component selects documents from vector databases, document repositories, or other knowledge stores and provides them as evidence to the generator. If this external knowledge source is manipulated, compromised, or poisoned, the model may incorporate false or misleading evidence into its reasoning process, producing outputs that appear coherent and well-supported while actually reflecting adversarial influence \cite{Chang_2025}. This risk is particularly severe in high-stakes domains such as the medical and healthcare field, where generated misinformation can lead to harmful outcomes \cite{Wang2024}.

Building on this context, this paper investigates how different \gls{RAG} design choices influence susceptibility to poisoning attacks. Unlike prior work that examines such factors individually, we conduct a full factorial experimental study that enables the quantification of higher-order interaction effects among pipeline components. We systematically vary a set of core pipeline parameters, including the dataset, retriever type, retrieval depth (top-$k$), chunk size, and chunk overlap, and the \gls{LLMs} used for generation. Crucially, we also introduce the number of databases and number of poisoned databases as experimental variables, a configuration dimension that has received little attention in prior work. To evaluate the impact of these factors, we assess the system under a new poisoning attack and measure retrieval and generation behavior through metrics such as Poison@k, Poison Rank, Score Margin, Abstention Rate, \gls{ASR}, and Correct Answer Rate. This experimental design makes it possible to analyze not only whether poisoning succeeds, but also under which compound conditions it becomes more effective, more stealthy, or more damaging to answer quality.

The remainder of this paper is organized as follows. Section~\ref{sec:2} reviews the relevant literature on \gls{RAG}, poisoning attacks. Section~\ref{sec:3} describes the experimental setup, including the selected datasets, retrieval models, poisoning strategy, parameter variations, and evaluation metrics. Section~\ref{sec:4} presents and discusses the obtained results, with a focus on identifying the factors that most strongly affect poisoning effectiveness. Finally, Section~\ref{sec:5} concludes the paper by summarizing the main findings, discussing their implications for secure \gls{RAG} design, and outlining directions for future work.

\section{Related Work}\label{sec:2}

Early work on \gls{RAG} primarily emphasized its benefits for factual grounding and access to external knowledge \cite{Lewis2020RetrievalAugmentedGF}. However, once researchers recognized that the generated response depends directly on retrieved evidence, the attention shifted toward the possibility that manipulating the knowledge source could systematically bias model outputs \cite{Zou_2025}. This shift established poisoning in \gls{RAG} not merely as a data-integrity problem, but as a pipeline-level vulnerability whose success depends on how retrieval and generation interact. The research on \gls{RAG} security then gradually evolved from demonstrating that poisoning attacks are possible to understanding the conditions under which they become effective \cite{zhang2026, Yanbo}.

Initial studies on poisoning in retrieval-augmented systems mostly focused on proving attack feasibility. These works showed that inserting or modifying malicious documents in the corpus can cause poisoned content to be retrieved and used as grounding evidence by the generator \cite{rayyan-375240496, rayyan-375241187}. Although these early studies were primarily attack-oriented, they already suggested the idea that poisoning effectiveness is strongly influenced by retrieval behavior. If poisoned documents are not retrieved, or are retrieved too low in the ranking, their impact on the final answer is often limited. Consequently, retrieval rank, similarity score, and inclusion in the final top-$k$ context emerged as early indicators of attack success.

As the field progressed, the literature began to identify more specific factors that affect poisoning outcomes. One important line of work studied how the structure and composition of the corpus influence vulnerability. These studies suggested that attack success depends not only on the presence of malicious documents, but also on how well they blend into the corpus distribution. Tan et al. \cite{rayyan-375822694} defines the challenge of attacking RAG systems in a dual nature objective. First, the injected content must be sufficiently relevant, according to the retriever's similarity function, to be consistently retrieved for a broad range of unseen user queries. Second, once retrieved, the same content must meaningfully influence the \gls{LLMs} generation process, potentially overriding safety mechanisms or factual grounding.

The literature also indicates that retrieval configuration itself is an important influence factor. Different retrievers do not expose poisoning in the same way, because sparse, dense, graph-based, and hybrid retrieval mechanisms rank evidence according to different principles \cite{karpukhin}. As a result, poisoning effectiveness may vary depending on the retriever architecture, the retrieval depth, and the scoring behavior used to construct the final context \cite{replug}. Similarly, the choice of top-$k$ affects how much opportunity poisoned documents have to appear in the retrieved set and whether they dominate or compete with clean evidence. These observations suggest that poisoning should not be analyzed independently of system configuration, since the same attack may behave differently under different retrieval strategies \cite{Zou_2025, replug}.

A related stream of work has examined how document preprocessing influences downstream robustness. Since most \gls{RAG} pipelines divide source documents into chunks before indexing, chunk size and chunk overlap can affect whether poisoned content remains concentrated, diluted, or repeatedly represented across the vector store \cite{abs-2312-10997}. Larger chunks may preserve more context and increase the semantic coherence of a poisoned passage, while smaller chunks may isolate the manipulated claim but also multiply its opportunities for retrieval if overlapping windows are used. Although these preprocessing choices are often treated as engineering details, the literature increasingly suggests that they can materially influence both retrieval behavior and poisoning exposure.

Beyond retrieval, recent studies also point to the generator as an influence factor. Different \gls{LLMs} vary in how strongly they rely on retrieved evidence, how easily they are persuaded by repeated claims, and how willing they are to abstain or reject unreliable context \cite{liang, dai2025}. Dai et al. \cite{dai2025} work shows that modern LLMs sometimes resist poisoned context through self-correction mechanisms, and attackers must suppress these behaviors to achieve reliable poisoning. This means that poisoning success is not determined only by whether malicious passages are retrieved, but also by how the generator interprets and prioritizes them relative to its internal knowledge and reasoning patterns. In this sense, poisoning in \gls{RAG} is a combined phenomenon shaped by both the exposure to the retriever-side and the susceptibility to the generator-side.

Overall, related work suggests that poisoning effectiveness in \gls{RAG} is shaped by a combination of interacting factors, including corpus realism, poisoning distribution, retrieval model, chunking strategy, and generator behavior. However, despite these insights, prior work has often studied such elements in isolation or in the context of specific attack demonstrations. Comparatively less attention has been given to systematically examining how varying core \gls{RAG} configuration parameters changes poisoning effectiveness across different settings. This gap motivates the present work, which investigates poisoning in \gls{RAG} through the perspective of influence factors rather than treating it only as a categorical security issue.

\section{Methodology}\label{sec:3}

This section describes the experimental methodology used to analyze poisoning behavior in RAG systems. The goal of the methodology is to systematically evaluate how different components of the RAG pipeline affect the success of poisoning attacks. To achieve this, we construct a controlled experimental setup in which individual pipeline components can be varied independently while measuring their impact on both retrieval behavior and generated responses.

\textbf{Dataset Selection}. The evaluation is conducted using two widely used question-answering datasets,  HotpotQA \cite{yang-etal-2018-hotpotqa} and MS-MARCO \cite{nguyen2016ms}. HotpotQA contains multi-hop reasoning questions that require integrating information from multiple documents, while MS MARCO focuses on open-domain question answering using web-scale passages. Together, these datasets capture complementary retrieval challenges, including multi-document reasoning and high-precision passage retrieval. To ensure an efficient evaluation, experiments are performed on a curated subset rather than on the full datasets. Specifically, we construct a subset of 100 semantically similar question pairs between HotpotQA and MS MARCO. This curated subset was chosen to enable a clearer cross-dataset comparison under aligned retrieval conditions, rather than to represent the full distribution of either dataset. It also reduces computational overhead while preserving diverse retrieval scenarios.

\textbf{Retrieval Architectures}. To examine how retrieval mechanisms influence poisoning exposure, the experiments evaluated three different retriever architectures that represent distinct ranking paradigms. BM25 is used as the representative sparse retriever. It ranks documents using term frequency and inverse document frequency statistics and serves as a strong baseline for traditional information retrieval. Dense BGE represents dense vector retrieval. In this setting, queries and documents are encoded into dense embeddings, and similarity is computed in vector space to retrieve semantically related passages. Graph-based retrieval incorporates document relationships through a structured graph representation. Instead of ranking documents solely by similarity, this approach retrieves evidence by traversing connections between related documents.

\textbf{Retrieval Depth}. The number of documents retrieved for each query is controlled using the top-$k$ parameter, which determines how many passages are returned by the retriever and are included in the model context. The experiments evaluate three retrieval depths, 2, 3, and 5. Varying the depth of retrieval allows the study of how poisoned passages compete with clean evidence in the retrieved set. Lower values of $k$ can restrict the available context and may amplify the influence of a single malicious document, while higher values increase the diversity of retrieved evidence, but can also increase the probability that poisoned content appears in the context window.

\textbf{Knowledge Base Configuration}. To simulate different poisoning scenarios, the experimental setup constructs RAG systems using multiple document collections. The number of available knowledge bases is varied between 1 and 2 databases to simulate redundant retrieval environments. In the two-database configuration, the system replicates the same dataset across both sources, designating one as a safe repository. Within these collections, specific databases may contain adversarial content. The extent of this compromise is evaluated in two distinct configurations, single-source poisoning (1 database) and multi-source poisoning (2 databases). By varying these parameters, the study evaluates the performance trade-offs across three distinct risk profiles, a baseline single-source compromise (1 database poisoned), a dual-source redundant environment with a point of failure (2 databases, 1 poisoned), and a system-wide saturation scenario where all available knowledge bases are corrupted (2 databases, 2 poisoned).

\textbf{Document Preprocessing}. Before indexing, documents are segmented into smaller units to support efficient retrieval. The segmentation process is controlled through two pre-processing parameters, chunk size and chunk overlap. The experiments evaluate two chunk sizes, 512 and 768 tokens. These values control the amount of contextual information preserved within each indexed segment. Larger chunks may retain more semantic context, while smaller chunks provide more granular retrieval units. To account for the possibility that relevant information may span chunk boundaries, overlapping segmentation is also applied. Two overlap values are evaluated, 64 and 100 tokens. Chunk overlap introduces redundancy across indexed segments and can increase the likelihood that specific information appears in multiple retrieved passages. 

\textbf{Generator Models}. The final stage of the RAG pipeline is the generator model, which produces answers using the retrieved context. Two LLMs are evaluated in the experiments, llama-4-scout-17b-16e-instruct and openai-gpt-oss-120b. These models differ in scale and training characteristics, providing an opportunity to analyze how generator behavior influences susceptibility to poisoned evidence. By evaluating multiple generators, the experiments assess whether poisoning success depends not only on retrieval exposure but also on how the language model interprets and prioritizes retrieved information.

\textbf{Poisoning Strategy}. To evaluate the robustness of RAG systems against knowledge poisoning, adversarial documents are automatically generated and inserted into selected knowledge bases. For each query, the poisoning process first determines a target claim, which represents an incorrect answer that the poisoned document will promote. If the question contains explicit alternatives (``A or B''), the incorrect option is selected directly. Otherwise, a language model generates several plausible distractor answers, and a filtering and ranking procedure selects the most suitable candidate while avoiding duplicates or variations of the correct answer. After selecting the target claim, an adversarial passage is generated using structured prompts. The generated text begins with a sentence that includes both the original question and the adversarial claim to ensure strong lexical overlap with the query, increasing the likelihood that the passage will be retrieved. The paragraph is written in a neutral encyclopedic tone and repeats key terms from the question to reinforce retrieval relevance, while explicitly avoiding the correct answer. Validation checks ensure that the generated text satisfies these constraints before the poisoned document is added to the knowledge base for evaluation. 

\textbf{Evaluation Metrics}. To quantify the system's behavior under adversarial conditions, we evaluate a series of metrics categorized by retrieval exposure and generation outcomes. At the retrieval level, we measure Poison@k, which indicates whether a poisoned document appears within the top-$k$ retrieved results. We also report the Poison Rank, defined as the ranking position of the poisoned document among the retrieved passages, and the Score Margin, which measures the similarity score difference between the best poisoned document and the highest-ranked clean document. At the generation level, we evaluate the model’s final response under poisoned retrieval. The generator can produce three possible outcomes. It may abstain from answering when the retrieved evidence appears unreliable, measured by the Abstention Rate. It may generate the correct answer, captured by the Correct Answer Rate. Alternatively, the poisoned content may influence the response and lead to a manipulated answer, measured by the ASR.

\section{Results and Discussion}\label{sec:4}

The experimental design combines all evaluated parameter settings into a full factorial grid, resulting in 432 unique configurations. To begin the analysis, the main effects of the experimental factors are examined using a \gls{OLS} regression model for each evaluation metric. The strongest effects across configurations are summarized in Tables~\ref{tab:retrieval_main_effects} and~\ref{tab:generation_main_effects}. We first focus on the retrieval-level metrics in Table~\ref{tab:retrieval_main_effects}, highlighting the five factors with the largest coefficients.

\begin{table}[h!]
\centering
\caption{Top 5 strongest main effects for retrieval-level metrics.}
\label{tab:retrieval_main_effects}
\resizebox{\textwidth}{!}{%
\begin{tabular}{llllll}
\hline
\multicolumn{2}{c}{\textbf{Poison@k}} &
\multicolumn{2}{c}{\textbf{Poison Rank}} &
\multicolumn{2}{c}{\textbf{Score Margin}} \\
\hline
Factor & Coef. & Factor & Coef. & Factor & Coef. \\
\hline
C(retriever)[T.Graph] & -0.170 & C(num\_databases)[T.2] & 1.690 & C(retriever)[T.dense\_bge] & -0.402 \\
C(dataset)[T.MS-MARCO] & -0.103 & C(num\_poisoned\_databases)[T.2] & -1.578 & C(retriever)[T.Graph] & -0.317 \\
C(top\_k)[T.5] & 0.092 & C(top\_k)[T.5] & 0.816 & C(dataset)[T.MS-MARCO] & -0.114 \\
C(top\_k)[T.3] & 0.042 & C(retriever)[T.Graph] & 0.475 & C(top\_k)[T.5] & -0.035 \\
C(retriever)[T.dense\_bge] & -0.040 & C(dataset)[T.MS-MARCO] & 0.338 & C(num\_poisoned\_databases)[T.2] & 0.025 \\
\hline
\end{tabular}%
}
\end{table}

Looking first at Poison@k, the graph-based retriever yields the largest negative effect, indicating that poisoned documents are less likely to appear among the retrieved results under this retrieval architecture. A similar, although smaller, reduction is observed for the dense BGE retriever. Overall, within the evaluated  poisoning strategy, these findings suggest that semantic and structure-aware retrieval methods are less exposed to the tested adversarial passages than purely lexical retrieval approaches. Dataset choice also has a substantial impact, experiments conducted on MS-MARCO show a lower probability of retrieving poisoned documents than those on HotpotQA. In addition, retrieval depth appear as a key factor. Increasing the number of retrieved documents from $k=2$ to $k=3$ and $k=5$ significantly increases Poison@k, indicating that larger retrieval increases the risk of poisoned passages being included in the retrieved context.

For Poison Rank, the results reveal a more different pattern. Using two databases increases Poison Rank, suggesting greater robustness, likely because poisoned documents face stronger competition from clean evidence. By contrast, increasing the number of poisoned databases decreases Poison Rank. Adding more poisoned databases increases the number of adversarial candidates available to the retriever. This raises the chance that at least one poisoned passage closely matches the query and achieves a high retrieval score. In line with the previous findings, retrieval architecture, retrieval depth, and dataset choice also affect how competitive poisoned passages are within the retrieved set. 

The Score Margin results show a pattern consistent with the previous findings, further highlighting the importance of the retrieval method. Both the dense BGE and graph-based retrievers reduce the Score Margin between poisoned and clean documents, indicating that adversarial passages are less competitive relative to legitimate evidence under these retrieval strategies. Similarly, MS-MARCO yields lower Score Margins than HotpotQA, reinforcing the conclusion that dataset characteristics play an important role in poisoning susceptibility.

We next turn to the generation-level metrics. Table~\ref{tab:generation_main_effects} summarizes the strongest main effects, listing for each metric the factors with the largest absolute coefficients, and thus those with the greatest influence on system behavior.

\begin{table}[h!]
\centering
\caption{Top 5 strongest main effects for generation-level metrics.}
\label{tab:generation_main_effects}
\resizebox{\textwidth}{!}{%
\begin{tabular}{llllll}
\hline
\multicolumn{2}{c}{\textbf{Abstention Rate}} &
\multicolumn{2}{c}{\textbf{Correct Answer Rate}} &
\multicolumn{2}{c}{\textbf{ASR}} \\
\hline
Factor & Coef. & Factor & Coef. & Factor & Coef. \\
\hline
C(llm)[T.openai-gpt-oss-120b] & -0.080 & C(retriever)[T.dense\_bge] & 0.098 & C(llm)[T.openai-gpt-oss-120b] & 0.169 \\
C(retriever)[T.Graph] & 0.078 & C(llm)[T.openai-gpt-oss-120b] & -0.088 & C(retriever)[T.Graph] & -0.143 \\
C(top\_k)[T.5] & -0.054 & C(num\_poisoned\_databases)[T.2] & -0.066 & C(retriever)[T.dense\_bge] & -0.073 \\
C(top\_k)[T.3] & -0.039 & C(retriever)[T.Graph] & 0.065 & C(num\_poisoned\_databases)[T.2] & 0.062 \\
C(retriever)[T.dense\_bge] & -0.025 & C(num\_databases)[T.2] & 0.060 & C(num\_databases)[T.2] & -0.044 \\
\hline
\end{tabular}%
}
\end{table}

For the Abstention Rate, the strongest effect is associated with the generator model openai-gpt-oss-120b that significantly decreases abstention, indicating that it is less likely to refuse answering when provided with retrieved context. Retrieval depth also contributes to this pattern, as increasing $k$ from 2 to 3 and 5 reduces abstention, suggesting that additional retrieved evidence increases the likelihood of the model answering. In contrast, the graph-based retriever increases abstention, which may reflect a tendency to promote more cautious behavior when the retrieved evidence is less straightforward to consolidate into a single response.

The results for the Correct Answer Rate further emphasize the importance of retrieval quality. Both the dense BGE and graph-based retrievers increase the probability of generating correct answers, suggesting that better retrieval directly improves the quality of the evidence available for generation. By contrast, openai-gpt-oss-120b decreases the Correct Answer Rate, indicating greater susceptibility to producing incorrect answers despite the retrieved context. The distribution of adversarial content also matters, increasing the number of poisoned databases reduces the Correct Answer Rate, whereas increasing the total number of databases improves it. This pattern suggests that generation quality depends on the balance between poisoned and clean evidence competing in the retrieval pool.

The ASR follows the same overall structure. The openai-gpt-oss-120b model has the largest positive effect on ASR, indicating greater vulnerability to poisoning attacks once adversarial content reaches the generator. In contrast, both the dense BGE and graph-based retrievers substantially reduce ASR under the evaluated attack, reinforcing that retrieval architecture is a key factor in this experimental setting. Consistent with the Correct Answer Rate results, increasing the number of poisoned databases increases ASR, while increasing the total number of databases reduces it. Taken together, these findings show that answer generation robustness is shaped by both model behavior and retrieval quality, with stronger retrievers not only improving answer correctness but also limiting the success of adversarial context injection. These results motivate pre-generation mitigation strategies that filter or down-rank suspicious retrieved passages before they can influence the LLM output, aligning with recent work on defensive filtering against RAG poisoning \cite{edemacu}.

While the previous analysis focused on the main effects of individual configuration parameters, poisoning behavior in \gls{RAG} systems may also emerge from interactions between multiple components of the pipeline. To capture these dependencies, an additional \gls{OLS} regression model was estimated including interaction terms. The model incorporates interactions up to the third order among all configuration variables. This specification allows the analysis to capture not only pairwise relationships but also higher-order interactions in which the combined effect of three parameters differs from the sum of their individual contributions. 

After fitting the model, the interaction coefficients were analyzed by grouping terms that involve the same set of parameters and computing the average absolute coefficient magnitude. This procedure provides an approximate measure of interaction strength and makes it possible to identify the parameter combinations with the largest influence on ASR. We focus on ASR because it is the most direct end-to-end measure of poisoning effectiveness \cite{Zou_2025}. The most influential interaction was \textit{dataset $\times$ retriever} (0.1551), followed by the \textit{dataset $\times$ retriever $\times$ top\_k} (0.1340). Other prominent effects included \textit{dataset $\times$ llm $\times$ retriever} (0.0769), \textit{dataset $\times$ llm} (0.0699), and \textit{retriever $\times$ top\_k} (0.0558). Together, these results justify focusing on \textit{dataset}, \textit{retriever}, \textit{top\_k} and \textit{llm} in the interaction analysis, as they appear most consistently in the strongest ASR-related interaction terms.

\begin{figure}[h!]\vspace*{4pt}
\centering
\includegraphics[width=0.83\textwidth]{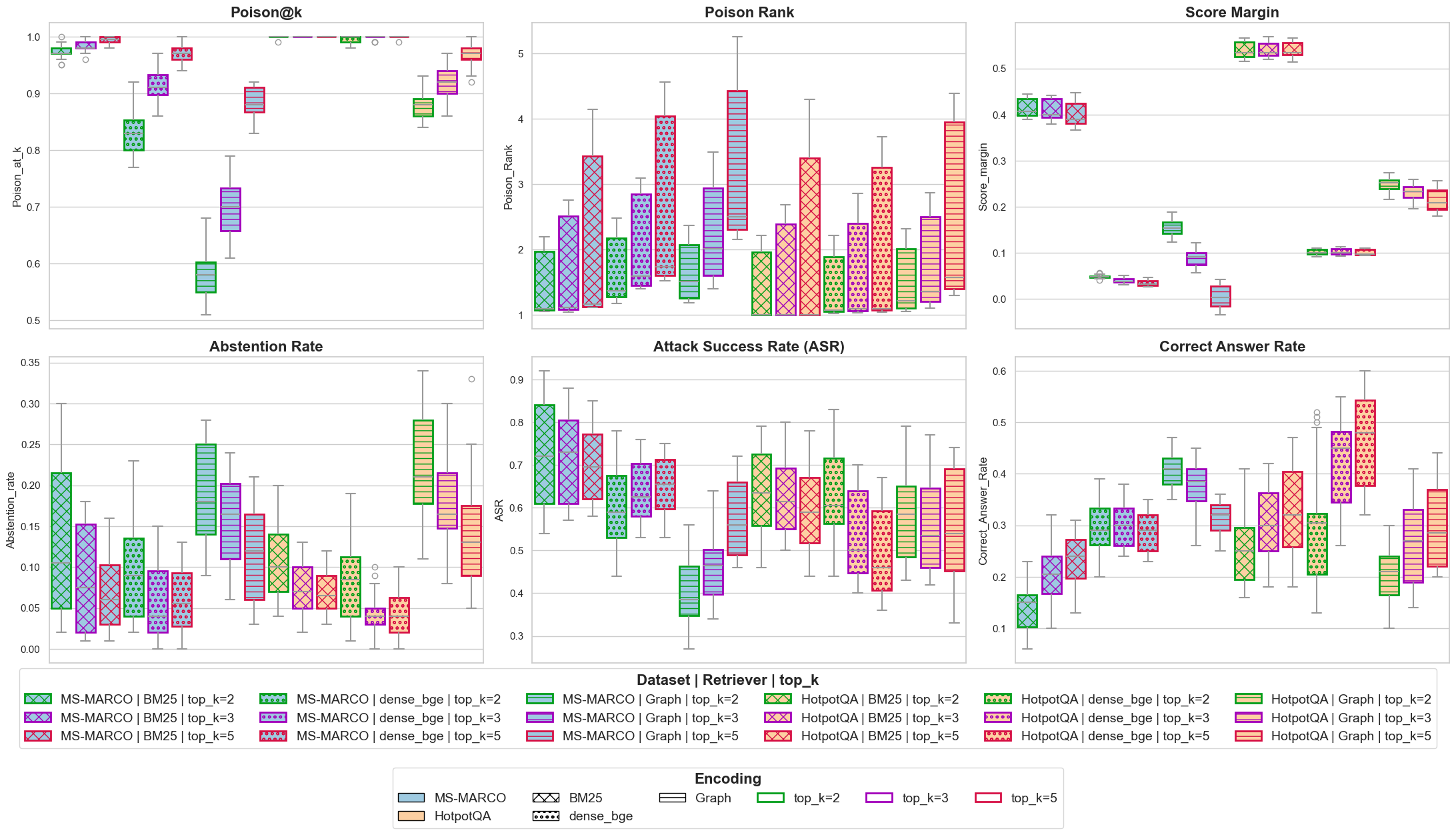}
\caption{Impact of the interaction between dataset, retriever, and retrieval depth on retrieval-level and generation-level metrics.}
    \label{fig:dataset_retriever_topk}
\end{figure}

Fig.~\ref{fig:dataset_retriever_topk} illustrates the combined effect of \textit{dataset}, \textit{retriever}, and \textit{top\_k} on all retrieval and generation metrics. Overall, the figure highlights that this interaction is one of the strongest in the model, particularly because the effect of increasing retrieval depth depends strongly on both the dataset and the retrieval architecture. For Poison@k, exposure is consistently highest for BM25 and generally lower for the dense BGE and graph-based retrievers, with the Graph retriever showing the largest reduction on MS-MARCO. At the same time, increasing $k$ tends to increase Poison@k across most settings, indicating that retrieving more documents raises the chance that poisoned passages appear in the retrieved set. This trend is especially pronounced for HotpotQA, where poisoning exposure remains high across retrievers and becomes nearly saturated at larger values of $k$.

The Poison Rank and Score Margin panels further clarify these retrieval effects. Poisoned documents tend to appear closer to the top of the ranking under BM25, especially on HotpotQA, whereas dense BGE and graph retrieval generally push poisoned passages lower. Similarly, Score Margin is substantially smaller for dense BGE and graph retrieval on MS-MARCO, showing that poisoned passages are less competitive relative to clean evidence in those settings. By contrast, HotpotQA exhibits markedly larger Score Margins, particularly under graph retrieval, suggesting that once poisoned documents are retrieved in this dataset, they remain comparatively strong competitors against legitimate passages. Together, these results indicate that the benefit of more robust retrievers is dataset-dependent, they are more effective on MS-MARCO than on HotpotQA.

The lower row of Fig.~\ref{fig:dataset_retriever_topk} shows how these retrieval differences propagate to generation outcomes. In particular, the Graph retriever is associated with higher Abstention Rate, especially on HotpotQA, suggesting that this retrieval strategy may induce more cautious model behavior when the evidence is difficult to reconcile into a confident answer. At the same time, dense BGE and graph retrieval achieve higher Correct Answer Rate than BM25 in most settings, particularly on MS-MARCO, indicating that improved retrieval quality generally benefits on final answer generation. The ASR mirrors the retrieval-level trends, ASR is highest for BM25, lower for dense BGE, and often lowest for the Graph retriever on MS-MARCO, while HotpotQA remains more vulnerable overall. Increasing $k$ frequently raises ASR under weaker retrieval settings, consistent with the higher poisoning exposure observed in the top row.

Taken together, these results shows that the effect of retrieval depth cannot be interpreted in isolation. Its impact depends strongly on the interaction between dataset and retriever, larger $k$ generally increases poisoning exposure, but the answer generation consequences are mitigated when retrieval is more robust and when the dataset provides stronger competing clean evidence. This helps explain why the \textit{dataset} $\times$ \textit{retriever} and \textit{dataset} $\times$ \textit{retriever} $\times$ \textit{top\_k} interactions emerge among the strongest effects in the regression analysis.

\begin{figure}[h!]\vspace*{4pt}
\centering
\includegraphics[width=0.83\textwidth]{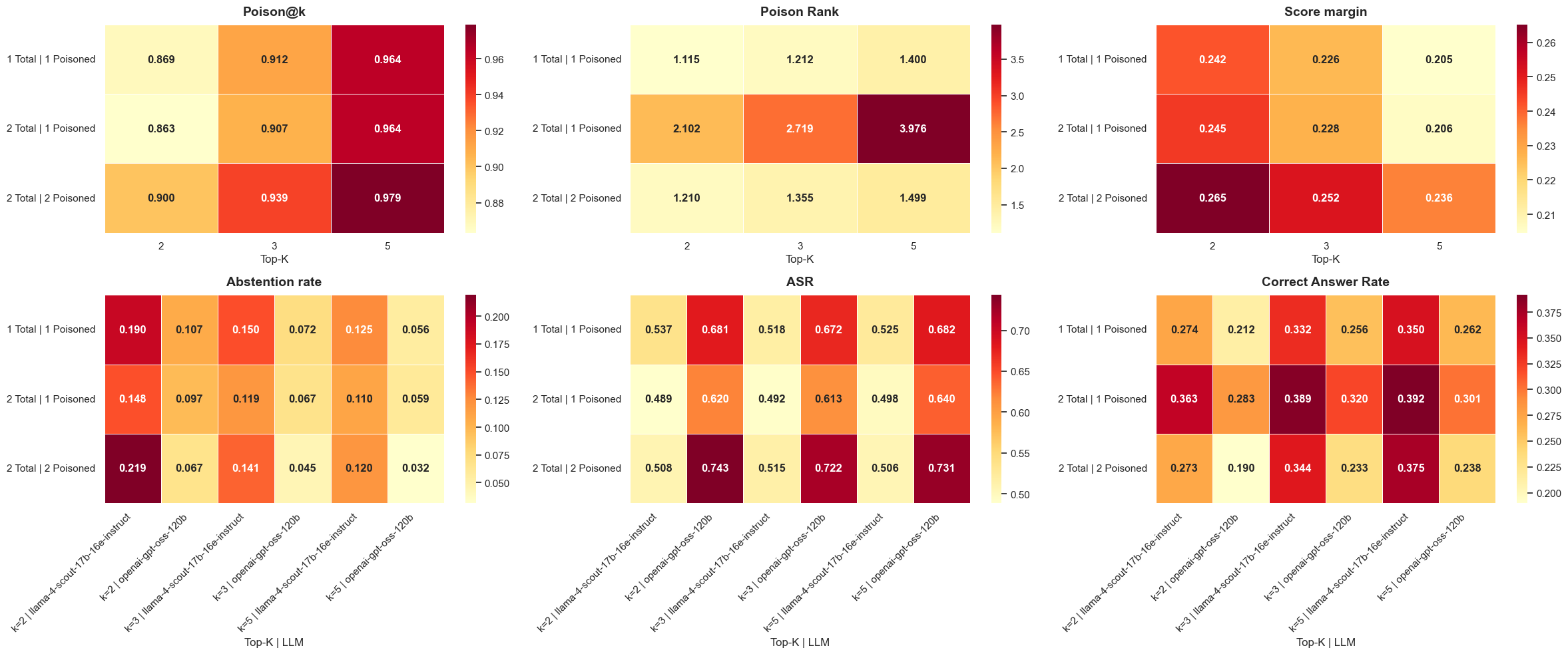}
    \caption{Mean retrieval and generation metrics across database configurations. The top row shows retrieval metrics as a function of database composition and retrieval depth ($k$), while the bottom row shows generation metrics additionally separated by generator model.}
    \label{fig:db_config_llm_heatmaps}
\end{figure}

Fig.~\ref{fig:db_config_llm_heatmaps} reports mean metric values across database configurations, retrieval depth, and generator model. Unlike the previous figure, which focused on retrieval-side factors, this plot explicitly introduces the LLM as an additional dimension in order to capture generation-stage variability. This is important because, even under the same retrieval conditions, different models may vary in abstention, correctness, and susceptibility to misleading evidence. The figure also examines a setting that is not commonly addressed in prior work, the distribution of poisoned content across multiple databases. By distinguishing among \textit{``1 Total $|$ 1 Poisoned''}, \textit{``2 Databases $|$ 1 Poisoned''}, and \textit{``2 Databases $|$ 2 Poisoned''}, the analysis goes beyond a simple poisoned-versus-clean comparison and shows how adversarial redundancy and competition from clean sources affect downstream attack success.

By analyzing the data in Fig.~\ref{fig:db_config_llm_heatmaps}, we confirm the patterns observed in the regression analysis at the retrieval level. In particular, Poison@k remains high across all database configurations and increases with retrieval depth, reinforcing the earlier result that broader retrieval increases poisoning exposure. Beyond this general trend, the figure shows that database composition mainly affects how strongly poisoned passages compete with clean evidence once they are retrieved. The \textit{``2 Databases $|$ 2 Poisoned''} configuration produces the highest Score Margin, indicating that poisoned passages remain more competitive when adversarial content is replicated across multiple databases. By contrast, the \textit{``2 Databases $|$ 1 Poisoned''} setting yields the highest Poison Rank, suggesting that the addition of a clean database pushes poisoned passages lower in the ranking. Together, these results indicate that adversarial redundancy and clean-source competition have opposite effects on retrieval robustness.

The generation-level metrics reveal a similarly clear effect of both model choice and database configuration. Across all settings, openai-gpt-oss-120b consistently exhibits lower Abstention Rate than llama-4-scout-17b-16e-instruct, indicating that it is less likely to withhold an answer when presented with retrieved evidence. However, this greater tendency to answer is accompanied by higher ASR, showing that the model is also more likely to follow poisoned context once adversarial passages are retrieved. In contrast, llama-4-scout-17b-16e-instruct tends to abstain more often, but this behavior is associated with lower attack success, suggesting a comparatively more cautious response to potentially misleading evidence.

Database composition further shapes these downstream outcomes. Moving from \textit{``2 Databases $|$ 1 Poisoned''} to \textit{``2 Databases $|$ 2 Poisoned''} generally increases ASR and reduces the Correct Answer Rate, indicating that replicating poisoned content across databases makes it more likely that misleading evidence will dominate the final generation context. Conversely, \textit{``2 Databases $|$ 1 Poisoned''} often achieves the highest Correct Answer Rate, suggesting that the presence of an additional clean database provides useful competing evidence that partially offsets the influence of poisoned passages.

\begin{figure}[h!]\vspace*{4pt}
\centering
\includegraphics[width=0.83\textwidth]{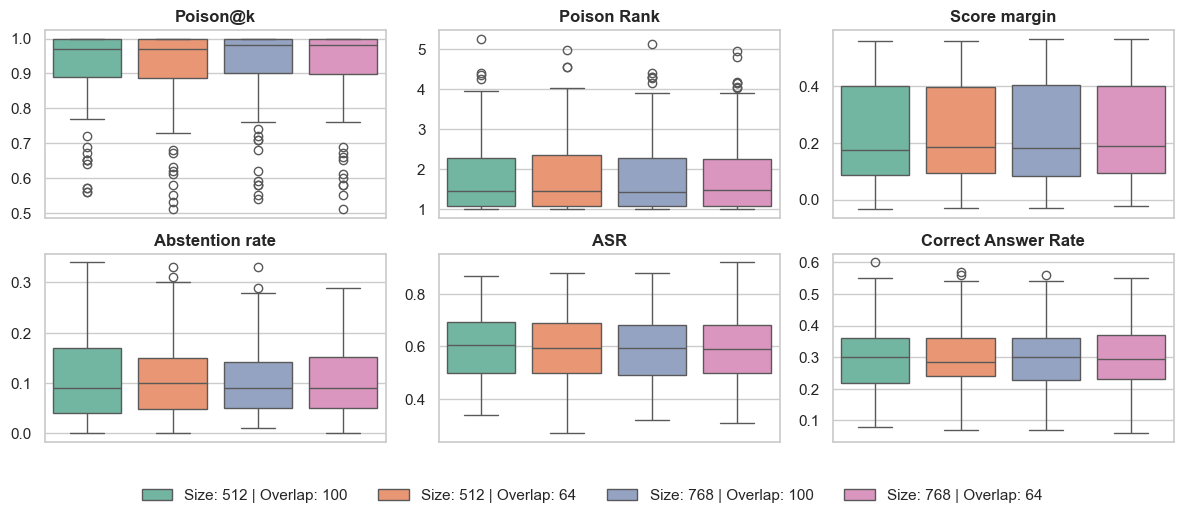}
\caption{Distribution of retrieval and generation metrics across different chunk size and overlap configurations.}
    \label{fig:Distribution}
\end{figure}

Finally, we also examined the effect of document chunking parameters, namely \textit{chunk size} and \textit{chunk overlap}, as shown in Fig.~\ref{fig:Distribution}. Across the evaluated configurations, these parameters exhibited only minor variations across both retrieval and generation metrics. In particular, differences between the tested settings (512 vs. 768 tokens and overlaps of 64 vs. 100) produced no consistent or substantial changes in Poison@k, Poison Rank, Score Margin, Abstention Rate, Correct Answer Rate, or ASR. This suggests that, within the examined ranges, chunking configuration has a comparatively limited influence on poisoning exposure and downstream generation behavior when compared to factors such as retriever architecture, dataset choice, or retrieval depth. For this reason, chunking parameters are not analyzed further in the remainder of the study.

\section{Conclusions}\label{sec:5}
This paper examined how selected design choices in RAG systems may influence susceptibility to poisoning attacks. Using a full factorial experimental design over 432 configurations, we analyzed the effects of dataset, retriever architecture, retrieval depth, number of databases, number of poisoned databases, chunking strategy, and generator model on retrieval and generation metrics. The results show that poisoning effectiveness is not determined by a single component, but by the integrated behavior of the full RAG pipeline. At the retrieval level, retriever architecture, dataset choice, and retrieval depth emerged as the main factors of poisoning exposure. In particular, dense BGE and graph-based retrievers generally reduced the frequency and competitiveness of poisoned passages compared with BM25, while larger values of $k$ increased the likelihood that poisoned documents appeared in the retrieved context. At the generation level, both retriever quality and generator choice had a significant impact on output robustness. Stronger retrievers improved answer correctness and reduced attack success, whereas the choice of LLM influenced how the model responded once poisoned evidence was present in the context. The interaction analysis further showed that these effects cannot be understood in isolation, with the strongest interaction terms involving dataset, retriever, retrieval depth, and LLM.

An additional contribution of this work is the explicit analysis of database composition, including scenarios with multiple poisoned databases. The results show that replicating adversarial content across databases increases the competitiveness of poisoned passages and amplifies resulting attack success, whereas the presence of additional clean databases can partially mitigate this effect by providing stronger competing evidence. This highlights the importance of studying poisoning not only as a property of individual documents, but also as a function of how adversarial content is distributed across the knowledge sources available to the system. Overall, the findings suggest that improving poisoning robustness in RAG requires a system-level perspective. Secure design choices should consider not only the retriever in isolation, but also how retrieval configuration, database composition, and generator behavior interact to shape final model outputs.

Future work should extend this analysis to larger and less curated query sets, additional datasets and generator models, and a broader range of poisoning and retrieval-manipulation attacks, including more stealthy and adaptive attack strategies. It would also be valuable to explore a wider set of configuration parameters, such as alternative chunking schemes, re-ranking stages, hybrid retrieval pipelines, and larger numbers of databases. Finally, further research should investigate how the susceptibility of different LLMs depends on their context window, since the amount of retrieved evidence that can be incorporated into the prompt may substantially affect both robustness and attack success.

\section*{Acknowledgements}
This work was supported by the AIDA project, which has received funding from the European Defence Fund (EDF) under grant agreement 101168202. This work has also received funding from UID/00760/2025.

\bibliography{sections/refs}
\bibliographystyle{elsarticle-harv}

\end{document}